\documentclass[a4paper,11pt]{article}
\pdfoutput=1 

\usepackage{jheppub} 

\usepackage[T1]{fontenc} 
\usepackage{graphicx}
\usepackage{amsfonts}
\usepackage{amssymb}
\usepackage{amsbsy}
\usepackage{amsmath}
\usepackage{mathrsfs}
\usepackage{latexsym}
\usepackage{natbib}
\usepackage{bm}
\usepackage{color}
\usepackage{braket}
\usepackage{slashed}
\usepackage{pgfplots}
\usepackage{caption}
\DeclareMathOperator{\sech}{sech}

\title{Characteristics of Z(2) multi-kink soliton configurations}

\author[1]{Susobhan Mandal}

\affiliation[1]{Department of Physical Sciences,\\ 
Indian Institute of Science Education and Research Kolkata,\\
Mohanpur - 741 246, WB, India }

\emailAdd{sm17rs045@iiserkol.ac.in}

\abstract{In this article, we first briefly review the solution of Z(2) kink solitons. Then 
we construct some multi-kink soliton configurations which are static and show their few 
features which are actually important to characterize their stability conditions. Not only
that this show also the particle characteristics of these kink configurations in these 
solitonic configurations. Then we will talk about dynamical kinks and show the affect of 
dynamics in the expression of force exerted by the neighbouring kink and anti-kink on each 
other in the multi-kink configurations. We have also defined an algebra through which we 
can write down equivalent ways of writing down multi-kink configuration mathematically.}

\begin{document} 
\maketitle
\flushbottom

\section{Introduction}
The soliton solutions are important in the studying models arising from various natural 
phenomena like the wave phenomena observed in fluid mechanics, optical fibers 
\cite{Frantzeskakis:97}, nuclear physics, high-energy physics, plasma physics, biology, 
solid-state physics, chemical kinematics etc.

Study of non-trivial field configurations in Quantum field theories is one of important aspects
of modern theoretical physics for different reasons. Convention quantum field theory(QFT) which
we are familiar with, assume the fact classical field around which study the excitations are 
spacetime independent. Though this is true for the bulk of the effect in QFT, but there are in 
fact states and effects which are due to non-trivial topologies of unperturbed state. And the
most important feature of these effects is that they can't be extracted using perturbative 
analysis \cite{sterman2001perturbative}. We begin by studying these configurations in Minkowski 
spacetime which are not excitations around a trivial classical configurations and these particles 
are collectively known as solitons.

A soliton is a non-dissipative non-trivial finite energy solution to field equation 
\cite{Nair:2005iw} coming from minimizing the action which describes the dynamics of the system. 
There is a subset, known as kinks which are also non dissipative non-trivial finite energy 
solutions \cite{Rajaraman:1982is}. What distinguishes the kinks from the other soliton 
configurations is that solitons remain unperturbed in collisions with other solitons, while this 
is not the case for kinks. A necessary condition for the existence of soliton and kink solutions 
to a field equation is that the potential energy in the action must have atleast 2 degenerate 
minimas. This is a consequence of the bounded energy of the kink.

In this article we will talk about $\mathbb{Z}(2)$ kink which arises from a $\phi^{4}$ like theory
and which has its own importance in physics for example nonlinear conductivity in some quasi-one-dimensional metals \cite{kimball1980kinks}, cooperative diffusion in one dimensional ionic 
conductors \cite{remoissenet1978cooperative}, spin-density excitations in 1-D organic polymers \cite{aslangul1983non}, nonlinear excitations in strongly-coupled plasma \cite{kourakis2006nonlinear},
existence of solitons in carbon nanotubes \cite{chamon2000solitons}. These are localized waves 
that propagate along one spatial direction with undeformed shape and some particle-like properties. 
We are mainly deal with time-independent kink solutions in scalar field theories with one spatial dimension since Derrick's theorem ruled out the existence of non-trivial time-independent finite 
energy kink soliton solutions in scalar field theories if the spatial dimension is greater than 1. 

\section{Scalar solitons in $1+1$ dimension}
\subsection{Brief review}
We start the discussion by mentioning classical scalar field theory in Minkowski spacetime. In conventional QFT, classical solutions are the nothing but vacuum configuration for which we 
quantize the theory. Thus understanding classical solution is necessary in order to understand 
full quantum theory. And to do that we basically study the solutions of Euler-Lagrange equation
\begin{equation}
\partial_{\mu}\frac{\partial\mathcal{L}}{\partial(\partial_{\mu}\phi)}=\frac{\partial\mathcal{L}}
{\partial\phi}
\end{equation}
coming from minimizing an action of following form
\begin{equation}
S[\phi]=\int d^{4}x\mathcal{L}(\phi(x),\partial\phi(x))
\end{equation}
As we will see, that equation (1) can have solution which varies over spacetime. In order for
these solutions to have physical importance, they must have finite energy given by following
expression
\begin{equation}
E=\int dx \mathcal{E}(x,t)
\end{equation}
where $\mathcal{E}=T_{ \ 0}^{0}(x,t)$ which is the (00)th component of the stress-energy tensor.
Furthermore, we want vacuum states which are stable under time evolution instead of turning again
into trivial configuration or solution at late times. This condition can be written as
\begin{equation}
\lim_{t\rightarrow\infty}\underset{\forall x}{\text{max}} \ \mathcal{E}(x,t)\neq0
\end{equation}
This is a physicist way of defining soliton which is different from its mathematical definition
which put further restriction on these class of solutions by imposing superposition principle.

The crucial aspect of all such solutions is going to be, what is the manifold of vacua for a 
theory. To understand this we started with a generic action in $1+1$ dimension given by
following Lagrangian density
\begin{equation}
\mathcal{L}=\frac{1}{2}\partial_{\mu}\phi\partial^{\mu}\phi-U(\phi)
\end{equation}
with $U(\phi)$ being some complicated function.

There could be many possible position where the potential is vanishing(even if it is non-zero 
minimum we can always make it zero). If we denote all such minimas with set $\{\phi_{i}\}$ then
$U(\phi_{i})=0, \ \forall i$.
The equation of motion is given by
\begin{equation}
\partial_{t}^{2}\phi-\partial_{x}^{2}\phi=-\frac{\partial U}{\partial\phi}
\end{equation}
and the energy density is given by
\begin{equation}
E[\phi]=\int_{-\infty}^{\infty}dx\Big[\frac{1}{2}(\partial_{t}\phi)^{2}+\frac{1}{2}(\partial_{x}
\phi)^{2}+U(\phi)\Big]
\end{equation}
Note that $E[\phi]=0$ means $\phi$ must be identically equal to one of the values in set 
$\{\phi_{i}\}$. But to have a non-zero finite energy solution al we want is that scalar field 
must approach one of the vacua $\{\phi_{i}\}$ asymptotically. Therefore, we can say that $E[\phi]
<\infty$ implies that
\begin{equation}
\lim_{x\rightarrow\infty}\phi(x)=\phi_{i}, \ \lim_{x\rightarrow-\infty}\phi=\phi_{j}
\end{equation}
So basically, finite energy solutions interpolate between two vacua or the zeros of the potential.

Ultimately we want to find explicitly the solutions of field equation which is rather difficult
in general but we can simplify our analysis by using Lorentz invariance. What we need to do is 
to first find static solutions and then boost it to find general time-dependent solutions.

\subsection{Static solution}
The field equation for static configurations become following
\begin{equation}
\partial_{x}^{2}\phi=\frac{\partial U}{\partial\phi}
\end{equation}
which is an ordinary second order differential equation.
Now multiply both side with $\partial_{x}\phi$ then we will get
\begin{equation}
\partial_{x}\phi\partial_{x}^{2}\phi=\partial_{x}\phi\frac{\partial U}{\partial\phi}\implies
\partial_{x}\Big[\frac{1}{2}(\partial_{x}\phi)^{2}-U(\phi)\Big]=0
\end{equation}
Therefore the quantity $W[\phi]=\frac{1}{2}(\partial_{x}\phi)^{2}-U(\phi)$ is independent of
spatial position and we know that at vacuas $\{\phi_{i}\}$, $W[\phi]=0$ which means we can say
that static configurations must satisfy
\begin{equation}
\frac{1}{2}(\partial_{x}\phi)^{2}-U(\phi)=0\implies\partial_{x}\phi=\pm\sqrt{2U(\phi)}
\end{equation}
Using these the above equation and previous facts about $\{\phi_{i}\}$ one can easily show 2
properties of these solutions which we state here without explicit proof
\begin{itemize}
\item In one spatial dimension, we can only have solutions that go between neighbouring vacua
which means solutions of eq.(2.11) can only connect $\phi_{i}$ and $\phi_{i+1}$ for any value
of $i$.
\item Any solution with same initial and final vacuum is a trivial solution.
\end{itemize}
Therefore, for potential $U(\phi)$ having $n$ number of minimas $\{\phi_{1},\ldots,\phi_{n}\}$
we can have possible transition which are the solutions
\begin{equation}
\phi_{1}\rightarrow\phi_{2},\ldots,\phi_{n-1}\rightarrow\phi_{n}, \ \ \phi_{n}\rightarrow\phi
_{n-1},\ldots,\phi_{2}\rightarrow\phi_{1}
\end{equation} 
Thus in total we have $2(n-1)$ possible solutions. The one goes in left order known as solitons 
and solutions in other directions are known as anti-solitons.

\subsection{$\mathbb{Z}(2)$ kinks}
To get an explicit solution we need to integrate out eq.(2.11) which will lead to following 
solutions for a general potential $U(\phi)$
\begin{equation}
x-x_{0}=\pm\int_{\phi_{0}}^{\phi(x)}\frac{d\tilde{\phi}}{\sqrt{2U(\tilde{\phi})}}
\end{equation}
We now consider kink solution which is a $1+1$ dimensional solution with potential $U(\phi)=
\frac{\lambda}{4}(\phi^{2}-\eta^{2})^{2}$. The higher dimensional version of these are domain
walls \cite{zvezdin2017dynamics}, \cite{semenoff2008domain}, \cite{krajewski2018domain}.

The vacua of this potential are at $\phi_{0}=\pm\eta$. And if we plug this expression into 
eq.(2.13) we will find
\begin{equation}
\begin{split}
x-x_{0} & =\pm\sqrt{\frac{2}{\lambda}}\int_{\phi_{0}}^{\phi(x)}d\tilde{\phi}(\tilde{\phi}^{2}
-\eta^{2})^{-1}\\
 & =\pm\frac{1}{\eta}\sqrt{\frac{2}{\lambda}}\tanh^{-1}\left(\frac{\phi(x)}{\eta}\right)\\
\implies\phi(x) & =\pm\eta\tanh(\sigma(x-x_{0})), \ \sigma=\eta\sqrt{\frac{\lambda}{2}} 
\end{split}
\end{equation}
This is final expression for $\mathbb{Z}(2)$ kink soliton solution where $+$ denote kink 
configuration and $-$ denote anti-kink configuration. We will come to why do we call it
$\mathbb{Z}(2)$ kink.

But before it we will show that the energy density of this configuration is localized in 
space. This determines the particle behaviour. The energy density of this configuration is
\begin{equation}
\mathcal{E}(x)=\frac{1}{2}(\partial_{x}\phi)^{2}+U(\phi)=\frac{\lambda\eta^{4}}{2}
\sech^{4}(\sigma(x-x_{0}))
\end{equation}
\begin{figure}
\centering
\includegraphics[height=5cm, width=7cm]{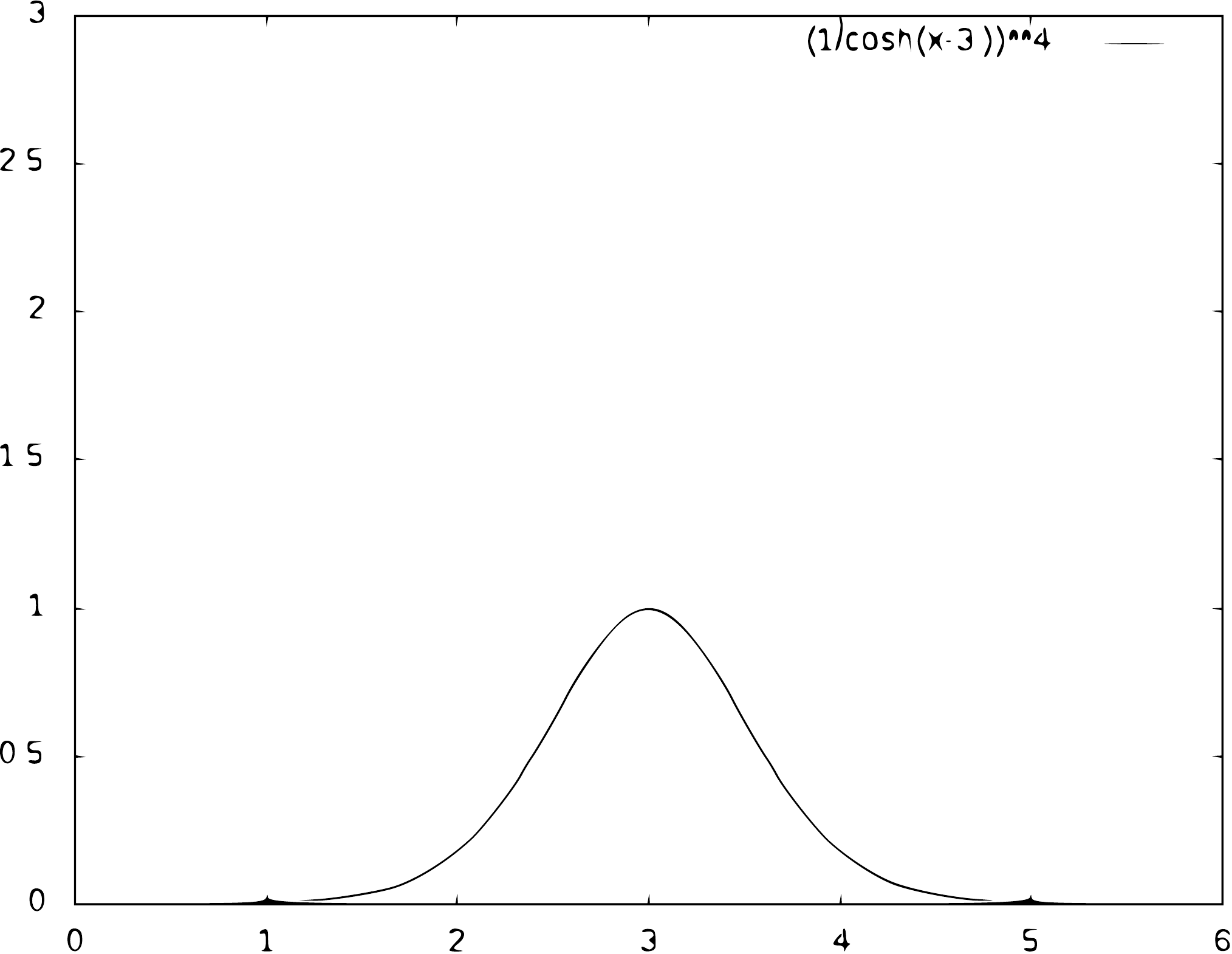}
\caption{plot of energy density for particular values of $\sigma,\eta$}
\end{figure}
And the total energy is 
\begin{equation}
E=\int_{-\infty}^{\infty}dx\mathcal{E}(x)=\frac{4}{3}\sigma\eta^{2}
\end{equation}
The reason we call it $\mathbb{Z}(2)$ kink solutions because the potential $U(\phi)$ has 
$\mathbb{Z}_{2}$ solution or in other-words $U(\phi)$ is invariant under $\phi\rightarrow
-\phi$ transformation. So, the vacua of this potential are related by $\mathbb{Z}_{2}$
action.

One may ask why the $\mathbb{Z}(2)$ kink cannot evolve into the trivial vacuum? For this to 
happen, the boundary condition at, say, $x=\infty$ would have to change in a continuous way 
from $\eta$ to $-\eta$. However, a small deviation of the field at infinity from one of the 
two vacua costs an infinite amount of potential energy. This is because as $\phi$ is changed, 
the field in an infinite region of space lies at a non-zero value of the potential. Hence, 
there is an infinite energy barrier to changing the boundary condition.  
To get time dependent solution we just need to boost transformation
\begin{equation}
x\mapsto\frac{x-vt}{\sqrt{1-v^{2}}}
\end{equation} 
then the solution becomes
\begin{equation}
\phi(x,t)=\eta\tanh(\sigma\gamma(x-x_{0}-vt))
\end{equation}
where $\gamma=\frac{1}{\sqrt{1-v^{2}}}$.

\subsection{Multi-kink configurations}
Since the kink solution is localize in space we can construct a field configuration with many 
such kink and anti-kink combination. However, we can arbitrarily combine them because of the 
constraint on $\mathbb{Z}(2)$ kink system is that kink must necessarily followed by an 
anti-kink since the asymptotic fields are restricted to vacuum $\phi_{0}=\pm\eta$. Therefore, 
it is not possible to have 2 neighbouring $\mathbb{Z}(2)$ kinks in any multi-kink configuration.

Now we describe what is known as "product ansatz" which is multi-kink configuration but not an
exact solution of eq.(2.11). This configuration can mathematically be written as product of 
alternative kink and anti-kink configurations in following way
\begin{equation}
\phi(x)=\frac{1}{\eta^{N+N'-1}}\prod_{i=1}^{N}\phi_{k}(x-k_{i})\prod_{j=1}^{N'}(-\phi_{k}
(x-l_{j}))
\end{equation} 
where $N,N'$ are the number of kinks and anti-kinks, $\phi_{k}$ is the kink solution. And 
$\{k_{i}\}$ and $\{l_{j}\}$ are the positions of kinks and anti-kinks such that $\ldots l_{i}
<k_{i}<l_{i+1}$. Note also that $|N-N'|\le1$ because kinks and anti-kinks must be in 
alternative position.
Product ansatz is generally a good approximation when as long as the positions of neighbouring 
kink and anti-kink are separated more than their widths. Therefore, in such case in the vicinity 
of $x=k_{i}$, only non-trivial factor is $\phi_{k}(x-k_{i})$ and rest of the factors are either
1 or -1.

Note that the factor $\frac{1}{\eta^{N+N'-1}}$ is introduced to maintain the dimensionality of 
the field variable $\phi(x)$ and also for normalization.

Another way to write an approximate multi-kink solution is addition approach. If $\phi_{i}$ 
denotes the $i^{\text{th}}$ kink(or anti-kink) in a sequence of N kinks and anti-kinks then
we can write following ansatz \cite{vachaspati2006kinks}
\begin{equation}
\phi(x)=\sum_{i=1}^{N}\phi_{i}(x)\pm(N_{2}-1)\eta, \ N_{2}=N(\text{mod} 2)
\end{equation} 
where $+$ is taken when leftmost is kink and $-$ is taken when leftmost is anti-kink.

Although the above ansatz are not exact solutions of the field equations in static limit but 
still they give field configurations which are made out of several widely spaced kinks and 
anti-kinks in smooth way. If we evolve this multi-kink configuration under field equation the
localized kinks and anti-kinks start moving due to forces exerted by other kinks and anti-kinks,
which we will discuss in next section. 

\section{Characteristics of multi-kink configurations}
\subsection{Multi-kink configuration in "Additive ansatz"}
We will follow \cite{vachaspati2006kinks} to calculate the amount of force through neighbouring
kink and anti-kink interact. To calculate that for simplicity we consider following configuration
where we consider one kink and one anti-kink configuration.
\begin{figure}
\centering
\includegraphics[height=5cm, width=7cm]{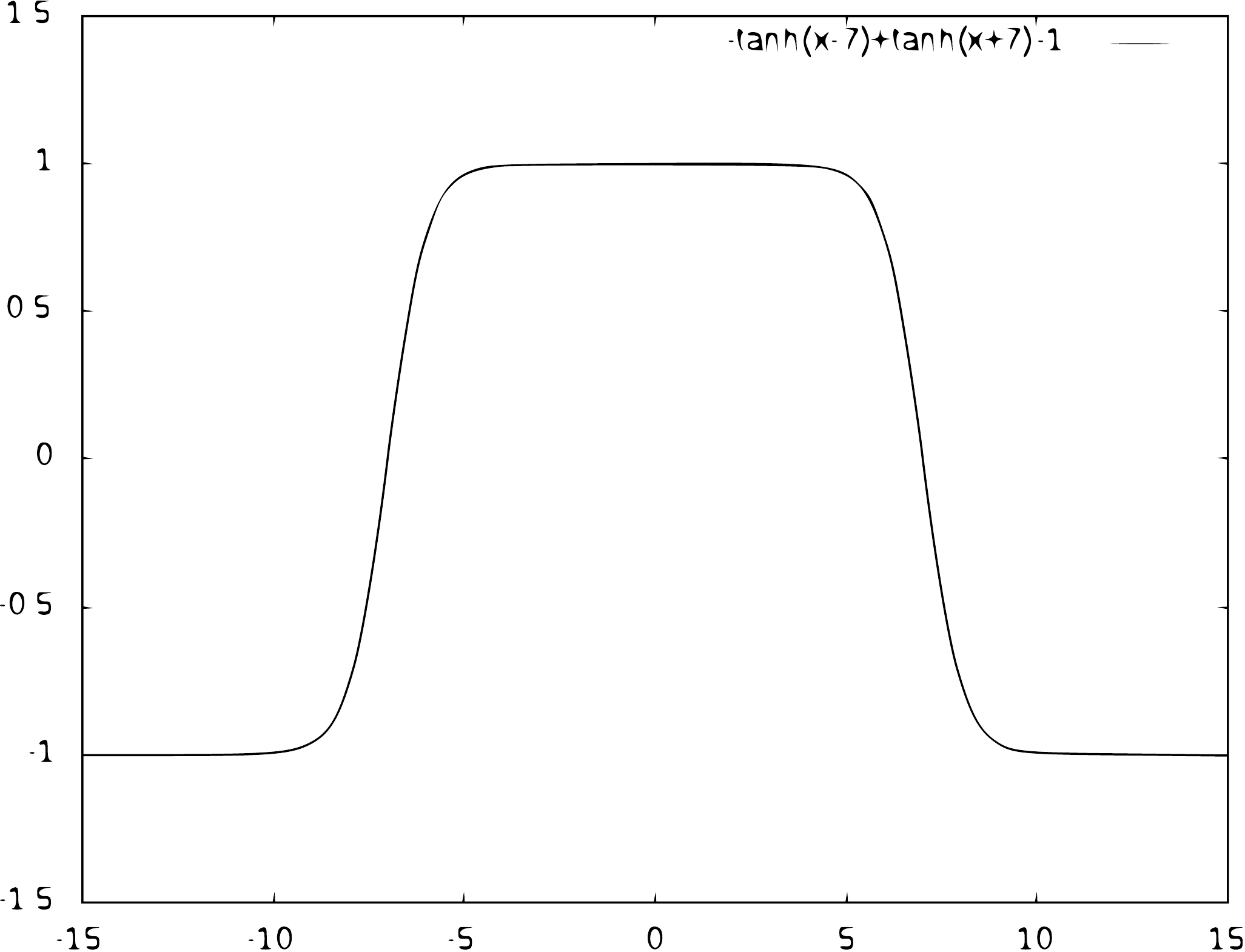}
\caption{kink, anti-kink combination in additive ansatz}
\end{figure}
Let's consider kink is located at a position $x=-a$ and anti-kink is centered at $x=a$ where
the separation is $2a$ which is larger than their individual width $2R$. Now we want to estimate 
the force on the kink owing to anti-kink.

To do that we will consider energy-momentum tensor, given by
\begin{equation}
T_{\mu\nu}=\partial_{\mu}\phi\partial_{\nu}\phi-\frac{1}{2}g_{\mu\nu}\Big[\frac{1}{2}
(\partial\phi)^{2}-V(\phi)\Big]
\end{equation}
where $g_{\mu\nu}=\text{diag}(1,-1)$ is the $1+1$ dimensional Minkowski metric.
The momentum of kink can be evaluated by integrating $T^{0i}=-T_{0i}$ from $-a-R$ to $-a+R$
in which kink is localized. Therefore, momentum of the kink is given by
\begin{equation}
\mathcal{P}=-\int_{-a-R}^{-a+R}dx\partial_{t}\phi\partial_{x}\phi
\end{equation}
Now we will use Newton's 2nd law to evaluate force and in the expression of time derivative of 
above equation we will use field equation
\begin{equation}
\begin{split}
F=\frac{d\mathcal{P}}{dt} & =-\int_{-a-R}^{-a+R}dx\Big[\partial_{t}^{2}\phi\partial_{x}\phi
+\partial_{t}\phi\partial_{t}\partial_{x}\phi\Big]\\
 & =-\int_{-a-R}^{-a+R}dx\Big[\partial_{x}^{2}\phi\partial_{x}\phi-\partial_{x}\phi V'(\phi)
 +\partial_{t}\phi\partial_{t}\partial_{x}\phi\Big]\\
 & =-\int_{-a-R}^{-a+R}dx\partial_{x}\Big[\frac{1}{2}(\partial_{x}\phi)^{2}+\frac{1}{2}
 (\partial_{t}\phi)^{2}-V(\phi)\Big]\\
 & =\Big[-\frac{1}{2}(\partial_{x}\phi)^{2}-\frac{1}{2}(\partial_{t}\phi)^{2}+V(\phi)\Big]_
{-a-R}^{-a+R} 
\end{split}
\end{equation}
To proceed further we will use the additive ansatz which in this case is following
\begin{equation}
\phi(t=0,x)=\phi_{k}(x)+\bar{\phi}_{k}(x)-\phi_{k}(\infty)=\phi_{k}(x)+\bar{\phi}_{k}(x)-
\eta
\end{equation}
In addition, we assume initially kinks are at rest which means $\partial_{t}\phi\Big|_{t=0}=0$.
And we will further use the fact that $(\partial_{x}\phi_{k})^{2}=2V(\phi_{k})$ which leads
to following expression
\begin{equation}
\begin{split}
F & =\Big[-\frac{1}{2}(\partial_{x}\phi)^{2}+V(\phi)\Big]_{-a-R}^{-a+R}\\
 & =\Big[-\partial_{x}\phi_{k}\partial_{x}\bar{\phi}_{k}+V(\phi_{k}+\bar{\phi}_{k}-\phi_{k}
(\infty))-V(\phi_{k})-V(\bar{\phi}_{k})\Big]_{-a-R}^{-a+R}
\end{split}
\end{equation}
We can safely ignore the derivative terms since we want to calculate the leading order term and
$\partial_{x}\phi_{k},\partial_{x}\bar{\phi}_{k}\rightarrow0$ near $x=-a$.
Now we define following quantities
\begin{equation}
\begin{split}
\phi_{k}^{\pm} & =\phi_{k}(-a\pm R), \ \bar{\phi}_{k}^{\pm}=\bar{\phi}_{k}(-a\pm R)\\
\Delta\phi_{k}^{\pm} & =\phi_{k}(-a\pm R)-\phi_{k}(\pm\infty)\\
\Delta\bar{\phi}_{k}^{\pm} & =\bar{\phi}_{k}(-a\pm R)-\bar{\phi}_{k}(-\infty)
\end{split}
\end{equation}
Now we also define $m_{\psi}^{2}=V''(\phi_{k}(\infty))=V''(\bar{\phi}_{k}(\infty))$.
After doing some simple algebraic manipulation we can write
\begin{equation}
F = m_{\psi}^{2}(\Delta\phi_{k}^{+}\Delta\bar{\phi}_{k}^{+}-\Delta\phi_{k}^{-}\Delta
\bar{\phi}_{k}^{-})
\end{equation}
Inserting the exact solutions $\phi_{k}(x)=\eta\tanh(\sigma(x+a))$ and $\bar{\phi}_{k}(x)=
-\eta\tanh(\sigma(x-a))$, Then we can write
\begin{equation}
\begin{split}
\Delta\phi_{k}^{+} & =\eta\tanh\sigma R-\eta\\
\Delta\phi_{k}^{-} & =-\eta\tanh\sigma R+\eta\\
\Delta\bar{\phi}_{k}^{+} & =\eta\tanh(\sigma(2a-R))-\eta\\
\Delta\bar{\phi}_{k}^{-} & =\eta\tanh(\sigma(2a+R))-\eta
\end{split}
\end{equation}
Then we can write
\begin{equation}
\begin{split}
F & =\eta^{2}m_{\psi}^{2}(1-\tanh\sigma R)[2-\tanh(\sigma(2a-R))-\tanh(\sigma(2a+R))]\\
 & \approx 4\eta^{2}m_{\psi}^{2}e^{-4\sigma a}\frac{1+e^{4\sigma R}}{1+e^{-2\sigma R}}
\end{split}
\end{equation}
which in leading order can be written as $F\approx 4\eta^{2}m_{\psi}^{2}e^{-4\sigma a}=
\frac{4 m_{\psi}^{4}}{\lambda}e^{-2a m_{\psi}}$.
The force is attractive since it is acting on the kink at $x=-a$ and points toward the anti-kink
at $x=a$. We will late discuss what does this expression physically mean?
\subsection{Multi-kink configuration in "Product ansatz"}
Note that in "product ansatz" all we need to is to put solution $\phi(x)=\frac{1}{\eta}
\phi_{k}(x)\bar{\phi}_{k}(x)$ therefore, only thing we need to change is the eq.(3.6) as 
follows
\begin{equation}
F=\Big[V\left(\frac{1}{\eta}\phi_{k}\bar{\phi}_{k}\right)-[V(\phi_{k})+V(\bar{\phi}_{k})]
\Big]_{-a-R}^{-a+R}
\end{equation}
which is easy to derive through field equation and algebraic manipulation.

In this case configuration schematically looks like following
\begin{figure}
\centering
\includegraphics[height=5cm, width=7cm]{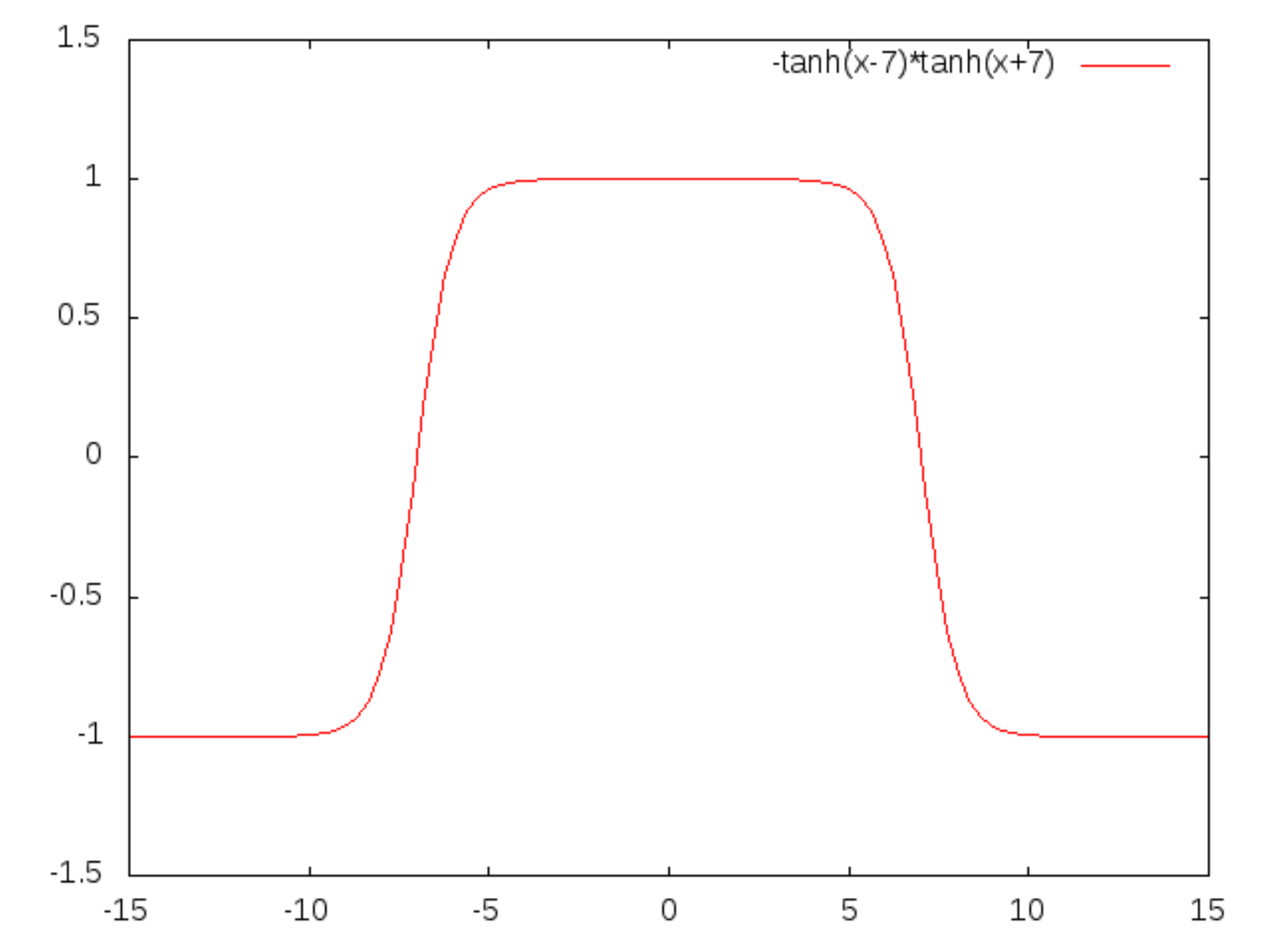}
\caption{kink, anti-kink combination in product ansatz}
\end{figure}
Now again we define following
\begin{equation}
\begin{split}
\phi_{k}^{\pm} & =\phi_{k}(-a\pm R), \ \bar{\phi}_{k}^{\pm}=\bar{\phi}_{k}(-a\pm R)\\
\Delta\phi_{k}^{\pm} & =\phi_{k}(-a\pm R)-\phi_{k}(\pm\infty)\\
\Delta\bar{\phi}_{k}^{\pm} & =\bar{\phi}_{k}(-a\pm R)-\bar{\phi}_{k}(-\infty)
\end{split}
\end{equation}
Then we can write eq.(3.11) as follows
\begin{equation}
\begin{split}
F & =\Big[V\left(\frac{1}{\eta}\phi_{k}^{+}\bar{\phi}_{k}^{+}\right)-V\left(\frac{1}{\eta}
\phi_{k}^{-}\bar{\phi}_{k}^{-}\right)\\
 & -[V(\phi_{k}^{+})+V(\bar{\phi}_{k}^{+})-V(\phi_{k}^{-})-V(\bar{\phi}_{k}^{-})]
\Big]\\
 & =\Big[V(\eta+\Delta\phi_{k}^{+}+\Delta\bar{\phi}_{k}^{+})-V(-\eta+\Delta\phi_{k}^{-}-\Delta
\bar{\phi}_{k}^{-})\\
 & -[V(\eta+\Delta\phi_{k}^{+})+V(\eta+\Delta\bar{\phi}_{k}^{+})-
V(-\eta+\Delta\phi_{k}^{-})-V(\eta+\Delta\bar{\phi}_{k}^{-})]\Big]\\
 & =V''(\eta)\Big[\Delta\phi_{k}^{+}\Delta\bar{\phi}_{k}^{+}+\Delta\phi_{k}^{-}\Delta
\bar{\phi}_{k}^{-}\Big]\\
 & =m_{\psi}^{2}(\Delta\phi_{k}^{+}\Delta\bar{\phi}_{k}^{+}+\Delta\phi_{k}^{-}\Delta
\bar{\phi}_{k}^{-})
\end{split}
\end{equation}
Note that to go from 2nd line to 3rd line we have used Taylor series expansion upto leading order 
which is 2nd order.
Again using the expressions for the solutions $\phi_{k}(x),\bar{\phi}_{k}(x)$ we can write
\begin{equation}
\begin{split}
F & =m_{\psi}^{2}\eta^{2}(\tanh\sigma R-1)[\tanh(\sigma(2a-R))-\tanh(\sigma(2a+R))]\\
 & \approx 4m_{\psi}^{2}\eta^{2}e^{-4\sigma a}\frac{e^{-\sigma R}}{e^{\sigma R}+e^{-\sigma R}}
 (e^{2\sigma R}-e^{-2\sigma R})\\
 & =4m_{\psi}^{2}\eta^{2}e^{-4\sigma a}e^{-\sigma R}(e^{\sigma R}-e^{-\sigma R})\\
 & =4m_{\psi}^{2}\eta^{2}e^{-4\sigma a}(1-e^{-2\sigma R})
\end{split}
\end{equation}
Note that although in leading order the expression of force matches with the answer we got in
previous subsection but higher order terms does not match. This is because both one kink and one 
anti-king with product and additive ansatz are almost similar with small changes which leads to
same leading order term but different higher order terms.

So, the magnitude of force in leading order in both cases is $F=\frac{4 m_{\psi}^{4}}{\lambda}
e^{-2a m_{\psi}}$.

\subsection{Physics behind expression of force}
The expression for the force could have been guessed from other considerations. The kinks are 
interacting through exchanging of massive scalar particles of mass $m_{\psi}$. As described in 
many QFT books \cite{Schwartz:2013pla}, \cite{Zee:2003mt}, \cite{Peskin:1995ev} the force 
mediated through scalar interactions is the Yukawa force which goes like $e^{-2m_{\psi}a}$. 

This is because of the fact if we want to study excitations near asymptotic vacua we can 
write $\phi=\eta+\psi$. Substituting this form in the Lagrangian (2.5) with potential mentioned 
before we can write action in following form
\begin{equation}
S=\int d^{2}x\Big[\frac{1}{2}\partial_{\mu}\psi\partial^{\mu}\psi-\frac{1}{2}m_{\psi}^{2}\psi^{2}
-\sqrt{\frac{\lambda}{2}}m_{\psi}\psi^{3}-\frac{\lambda}{4}\psi^{4}\Big]
\end{equation} 
which exactly justifies our answer.

\subsection{Multi-kink configuration in "product ansatz" for 2 kinks and 1 anti-kink}
In this subsection we will talk about configuration of following form
\begin{equation}
\phi=\phi_{k}^{(1)}\bar{\phi}_{k}^{(2)}\phi_{k}^{(3)}
\end{equation}
where 2 kinks are located at $x=-a,3a$ and anti-kink is located at $x=a$. Following is the
schematic diagram 
\begin{figure}
\centering
\includegraphics[height=5cm, width=9cm]{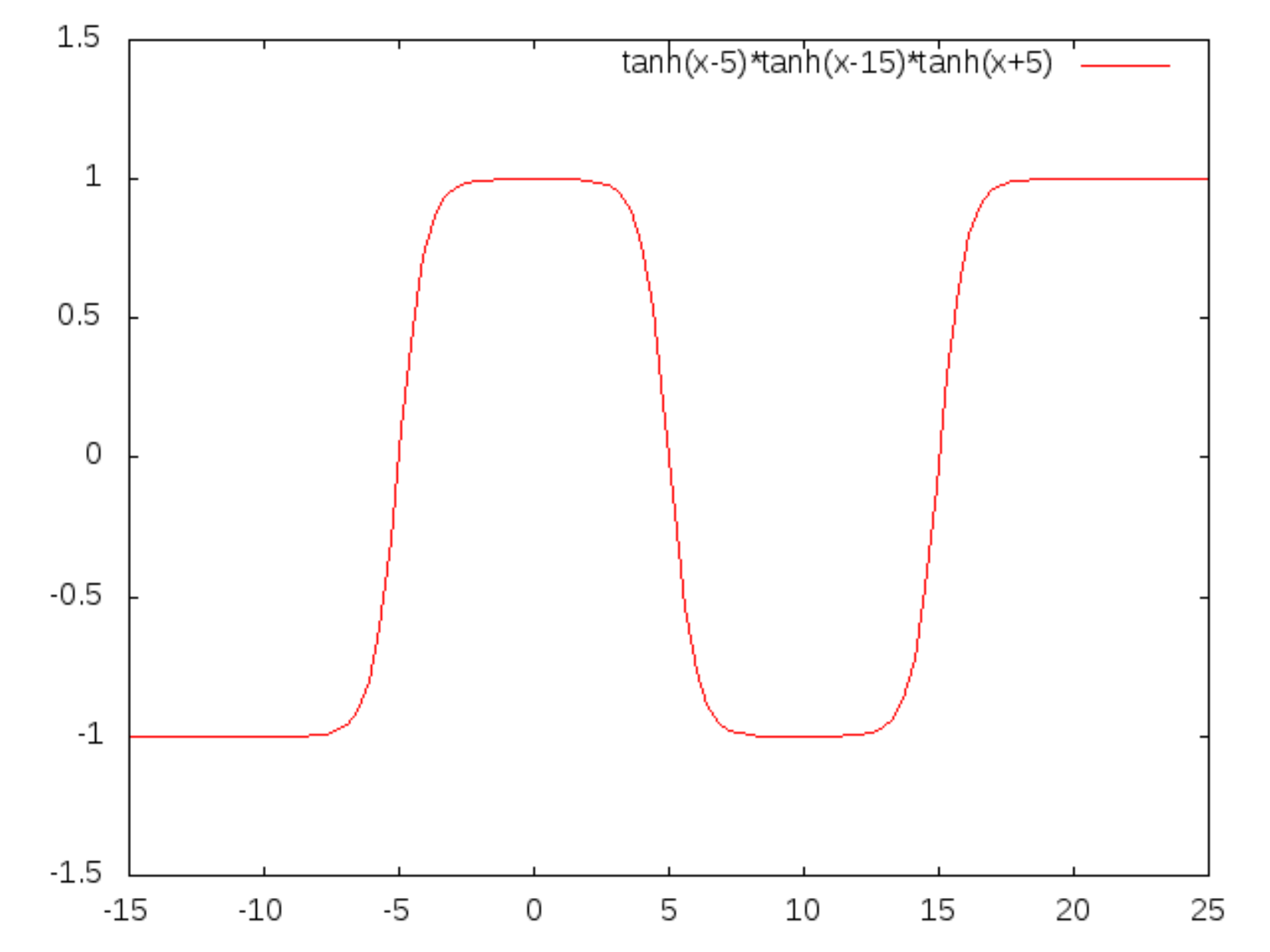}
\caption{product ansatz for 2 kinks and 1 anti-kink}
\end{figure}
Note that here we have following functions
\begin{equation}
\begin{split}
\phi_{k}^{(1)}(x) & =\eta\tanh(\sigma(x+a))\\
\bar{\phi}_{k}^{(2)}(x) & =-\eta\tanh(\sigma(x-a))\\
\phi_{k}^{(3)}(x) & =\eta\tanh(\sigma(x-3a))
\end{split}
\end{equation}
And we now define following quantities
\begin{equation}
\begin{split}
\phi_{k}^{(1)\pm} & =\phi_{k}^{(1)}(a\pm R)\\
\bar{\phi}_{k}^{(2)\pm} & =\bar{\phi}_{k}^{(2)}(a\pm R)\\
\phi_{k}^{(3)\pm} & =\phi_{k}^{(3)}(a\pm R)\\
\Delta\phi_{k}^{(1)\pm} & =\phi_{k}^{(1)}(a\pm R)-\eta\\
\Delta\bar{\phi}_{k}^{(2)\pm} & =\bar{\phi}_{k}^{(2)}(a\pm R)\pm\eta\\
\Delta\phi_{k}^{(3)\pm} & =\phi_{k}^{(3)}(a\pm R)+\eta
\end{split}
\end{equation}
In this subsection our aim is to measure the net force exerted on the anti-kink due to 2 kinks
on opposite side. According to Newton's law of point particle picture this should come out to
be zero but we need to check it rigorously. If this holds the particle characteristics of 
$\mathbb{Z}(2)$ kinks is proven.

Note that as earlier in this case, net force acting on anti-kink is given by
\begin{equation}
\begin{split}
F & =\Big[V\left(\frac{1}{\eta^{2}}\phi_{k}^{(1)}\bar{\phi}_{k}^{(2)}\phi_{k}^{(3)}\right)
-V(\phi_{k}^{(1)})-V(\bar{\phi}_{k}^{(2)})-V(\phi_{k}^{(3)})\Big]_{a-R}^{a+R}\\
 & =\Big[V\left(\frac{1}{\eta^{2}}\phi_{k}^{(1)+}\bar{\phi}_{k}^{(2)+}\phi_{k}^{(3)+}\right)
-V\left(\frac{1}{\eta^{2}}\phi_{k}^{(1)-}\bar{\phi}_{k}^{(2)-}\phi_{k}^{(3)-}\right)\\
 & -V(\phi_{k}^{(1)+})-V(\bar{\phi}_{k}^{(2)+})-V(\phi_{k}^{(3)+})
 +V(\phi_{k}^{(1)-})+V(\bar{\phi}_{k}^{(2)-})+V(\phi_{k}^{(3)-})\Big]\\
 & =\Big[V(\eta+\Delta\phi_{k}^{(1)+}-\Delta\phi_{k}^{(2)+}-\Delta\phi_{k}^{(3)+})
 - V(-\eta-\Delta\phi_{k}^{(1)-}-\Delta\phi_{k}^{(2)-}+\Delta\phi_{k}^{(3)-})\\
 & -V(\eta+\Delta\phi_{k}^{(1)+})-V(-\eta+\Delta\bar{\phi}_{k}^{(2)+})-V(-\eta+\Delta
\phi_{k}^{(3)+})+V(\eta+\Delta\phi_{k}^{(1)-})\\
 & +V(\eta+\Delta\bar{\phi}_{k}^{(2)-})+V(-\eta+\Delta\phi_{k}^{(3)-})\Big]
\end{split}
\end{equation} 
Now after doing Taylor series expansion upto 2nd order we will find
\begin{equation}
\begin{split}
F & =\frac{m_{\psi}^{2}}{2}\Big[(\Delta\phi_{k}^{(1)+}-\Delta\phi_{k}^{(2)+}-\Delta\phi_{k}
^{(3)+})^{2}-(-\Delta\phi_{k}^{(1)-}-\Delta\phi_{k}^{(2)-}+\Delta\phi_{k}^{(3)-})^{2}\\
 & -(\Delta\phi_{k}^{(1)+})^{2}-(\Delta\bar{\phi}_{k}^{(2)+})^{2}-(\Delta
\phi_{k}^{(3)+})^{2}+(\Delta\phi_{k}^{(1)-})^{2}+(\Delta\bar{\phi}_{k}^{(2)-})^{2}+(\Delta
\phi_{k}^{(3)-})^{2}\Big]\\
 & =m_{\psi}^{2}\Big[\Delta\bar{\phi}_{k}^{(2)+}\Delta\phi_{k}^{(3)+}+\Delta\bar{\phi}_{k}
^{(2)-}\Delta\phi_{k}^{(3)-}-\Delta\phi_{k}^{(1)+}\Delta\bar{\phi}_{k}^{(2)+}-\Delta\phi_{k}
^{(1)-}\Delta\bar{\phi}_{k}^{(2)-}\\
 & +\Delta\phi_{k}^{(1)-}\Delta\phi_{k}^{(3)-}-\Delta\phi_{k}^{(1)+}\Delta\phi_{k}^{(3)+}\Big]
\end{split}
\end{equation}
After doing simple algebra one would find
\begin{equation}
\begin{split}
\Delta\bar{\phi}_{k}^{(2)+}\Delta\phi_{k}^{(3)+} & +\Delta\bar{\phi}_{k}
^{(2)-}\Delta\phi_{k}^{(3)-}=4\eta^{2}e^{-4\sigma a}(1-e^{-2\sigma R})\\
\Delta\phi_{k}^{(1)+}\Delta\bar{\phi}_{k}^{(2)+} & +\Delta\phi_{k}
^{(1)-}\Delta\bar{\phi}_{k}^{(2)-}=4\eta^{2}e^{-4\sigma a}(1-e^{-2\sigma R})
\end{split}
\end{equation}
which means in line before last line of eq.(3.22) 4 terms cancelled each other. Therefore, only
thing left to check is following
\begin{equation}
\begin{split}
\Delta\phi_{k}^{(1)-}\Delta\phi_{k}^{(3)-} & -\Delta\phi_{k}^{(1)+}\Delta\phi_{k}^{(3)+}
=(\eta\tanh(\sigma(2a-R))\eta)(\eta-\eta\tanh(\sigma(2a+R)))\\
 & -(\eta\tanh(\sigma(2a+R))\eta)(\eta-\eta\tanh(\sigma(2a-R)))\\
 & =-\eta^{2}(1-\tanh(\sigma(2a-R)))(1-\tanh(\sigma(2a+R)))\\
 & +\eta^{2}(1-\tanh(\sigma(2a+R)))(1-\tanh(\sigma(2a-R)))\\
 & =0
\end{split}
\end{equation}
Therefore, upto 2nd order terms in Taylor series we found $F_{\text{net}}=0$ which shows particle
characteristic. But to make sure $F_{\text{net}}$ exactly equal to $0$, we need to go beyond 2nd
order in Taylor series expansion.

Now we do 3rd order where all we have to do is take the 3rd order term from the Taylor series
expansion of eq.(3.21) with using the fact that $V'''(-\eta)=-V'''(\eta)$. Then we can write 
at 3rd order net force to be following
\begin{equation}
\begin{split}
F & =\frac{V'''(\eta)}{3!}\Big[(\Delta\phi_{k}^{(1)+}-\Delta\bar{\phi}_{k}^{(2)+}-\Delta
\phi_{k}^{(3)+})^{3}+(\Delta\phi_{k}^{(3)-}\Delta\phi_{k}^{(1)-}-\Delta\bar{\phi}_{k}^{(2)
-})^{3}\\
 & -(\Delta\phi_{k}^{(1)+})^{3}+(\Delta\bar{\phi}_{k}^{(2)+})^{3}+(\Delta\phi_{k}^{(3)+})
^{3}+(\Delta\phi_{k}^{(1)-})^{3}+(\Delta\bar{\phi}_{k}^{(2)-})^{3}-(\Delta\phi_{k}^{(3)-})
^{3}\Big]\\
 & =\frac{V'''(\eta)}{3!}\Bigg[\Big[-3(\Delta\phi_{k}^{(1)+})^{2}\Delta\bar{\phi}_{k}^{(2)+}
-3(\Delta\phi_{k}^{(1)+})^{2}\Delta\phi_{k}^{(3)+}+3\Delta\phi_{k}^{(1)+}(\Delta\bar{\phi}
_{k}^{(2)+})^{2}\\
 & -3(\Delta\bar{\phi}_{k}^{(2)+})^{2}\Delta\phi_{k}^{(3)+}+3\Delta\phi_{k}^{(1)+}(\Delta
\phi_{k}^{(3)+})^{2}-3\Delta\bar{\phi}_{k}^{(2)+}(\Delta\phi_{k}^{(3)+})^{2}+6\Delta\phi_{k}
^{(1)+}\Delta\bar{\phi}_{k}^{(2)+}\Delta\phi_{k}^{(3)+}\Big]\\
 & +\Big[-3(\Delta\phi_{k}^{(3)-})^{2}\Delta\phi_{k}^{(1)-}-3(\Delta\phi_{k}^{(3)-})^{2}
\Delta\bar{\phi}_{k}^{(2)-}+3(\Delta\phi_{k}^{(1)-})^{2}\Delta\phi_{k}^{(3)-}-3(\Delta\phi
_{k}^{(1)-})^{2}\Delta\bar{\phi}_{k}^{(2)-}\\
 & +3\Delta\phi_{k}^{(3)-}(\Delta\bar{\phi}_{k}^{(2)-})
^{2}+6\Delta\phi_{k}^{(1)-}\Delta\bar{\phi}_{k}^{(2)-}\Delta\phi_{k}^{(3)-}\Big]\Bigg] 
\end{split}
\end{equation}
Now if we substitute following
\begin{equation}
\begin{split}
\Delta\phi_{k}^{(1)\pm} & =\eta(\tanh(\sigma(2a\pm R))-1)\\
\Delta\bar{\phi}_{k}^{(2)\pm} & =\pm\eta(1-\tanh\sigma R)\\
\Delta\phi_{k}^{(1)\pm} & =\eta(1-\tanh(\sigma(2a\mp R)))
\end{split}
\end{equation}
in eq.(3.25) we will find that this 3rd contribution also vanishes. Following similar approach 
one can show that even at 4th order there will be no contribution in force acting on anti-kink
due to 2 kinks. Therefore, $F_{\text{net}}=0$ truly which means kinks and anti-kinks both shows
particle like behaviour.

\subsection{General Multi-kink configuration}
Note that according to earlier facts, if we form a multi-kink configuration initially then all 
the kinks and ant-kinks in the bulk don't feel forces from neighbouring anti-kinks or kinks.
But the kinks or anti-kinks located at the boundary of this configuration feel attractive force 
towards bulk which ultimately leads to collision between neighbouring kink and anti-kink at 
boundary which deforms the configuration at the boundary locally. But such deformation 
propagates toward bulk and eventually destroy the whole configuration. Therefore, these 
multi-kink configurations which are static initially, form a dynamical instability and so they 
are unstable configurations.

\section{Dynamical Kink and Correction to force expression}
We now want to look at kinks that move. Given what we have done so far, this is trivial in a 
sense of considering static configurations. Our theory is Lorentz invariant, so we simply apply 
a Lorentz boost. This changes kink solution in following way
\begin{equation}
\phi_{k}(x)\rightarrow\phi_{k}(x,t)=\eta\tanh(\sigma\gamma(x-x_{0}-vt))
\end{equation}
where $\gamma=\frac{1}{\sqrt{1-v^{2}}}$ and under small velocity approximation $v\ll 1$, we can
write the above solution as
\begin{equation}
\phi_{k}(x,t)\approx\eta\tanh(\sigma(x-a(t))), \ a(t)=x_{0}+vt
\end{equation}
Note that this solution satisfy
\begin{equation}
\partial_{t}\phi_{k}=-\dot{a}(t)\partial_{x}\phi_{k}\implies |\partial_{t}\phi_{k}|\ll 1
\end{equation}
Although, the configuration (4.2) is not an exact solution of field equation but we can think
of it as moving king with small velocity $\dot{a}(t)$.

Now we go back to situation in subsection (3.2) with product ansatz but with dynamical kins
and want to measure force exerted on kink at a given time. We follow similar approach but only
modification comes as follows
\begin{equation}
\begin{split}
F & =\Big[\frac{1}{2}(\dot{a}^{2}+1)(\partial_{x}\phi)^{2}+V(\phi)\Big]_{-a-R}^{-a+R}\\
 & =\Big[V\left(\frac{1}{\eta}\phi_{k}\bar{\phi}_{k}\right)-[V(\phi_{k})+V(\bar{\phi}
_{k})](\dot{a}^{2}+1)\Big]_{-a-R}^{-a+R}
\end{split}
\end{equation}
Now here we want to point out the fact that we are able to carry out integral over $T_{0i}$ 
to evaluate momentum because we do it at a given instant of time.

After some simple algebra one can show that only new thing that we need to calculate and 
which contribute as a correction in force is following
\begin{equation}
\begin{split}
F_{\text{correction}} & =V''(\eta)\Big[\frac{\dot{a}^{2}}{2}[(\Delta\phi_{k}^{-})^{2}
+(\Delta\bar{\phi}_{k}^{-})^{2}-(\Delta\phi_{k}^{+})^{2}-(\Delta\bar{\phi}_{k}^{+})
^{2}]\Big]\\
 & =\frac{m_{\psi}^{2}\dot{a}^{2}\eta^{2}}{2}\Big[(1-\tanh\sigma R)^{2}+(1-\tanh(\sigma
 (2a+R)))^{2}\\
 & -(1-\tanh\sigma R)^{2}-(1-\tanh(\sigma(2a-R)))^{2}\Big]\\
 & =\frac{m_{\psi}^{2}\dot{a}^{2}\eta^{2}}{2}\Big[(1-\tanh(\sigma(2a+R)))^{2}-(1-\tanh(
\sigma(2a-R)))^{2}\Big]\\
 & \approx m_{\psi}^{2}\dot{a}^{2}\eta^{2}[\tanh(\sigma(2a-R))-\tanh(\sigma(2a+R))]\\
 & \approx -2m_{\psi}^{2}\dot{a}^{2}\eta^{2}e^{-4\sigma a}(e^{2\sigma R}-e^{-2\sigma R})
\end{split}
\end{equation}
Note that correction factor is small because $\dot{a}^{2}\ll 1$ compared to the leading order
expression for force.

\section{Algebra of $\mathbb{Z}(2)$ kinks and anti-kinks}
Let's denote kink and anti-kink solutions by $\phi_{k}$ and $\bar{\phi}_{k}$ as earlier but this
time we will use an extra level to denote the location of kinks and anti-kinks. So, we denote
by $\phi_{k}^{(a)}$ a kink located at $x=a$ position and $\bar{\phi}_{k}^{(-a)}$ an anti-kink
located at $x=-a$ position.
so if we consider distance between successive kink and anti-kink is $2a$. Then we can write down 
following things(assuming left most is kink and located at $x=-a$) for 2 and 3 localized states
\begin{equation}
\begin{split}
\frac{1}{\eta}\phi_{k}^{(-a)}\bar{\phi}_{k}^{(a)} & \equiv\phi_{k}^{(-a)}+\bar{\phi}_{k}
^{(a)}\\
\frac{1}{\eta}\phi_{k}^{(-a)}\bar{\phi}_{k}^{(a)}+\phi_{k}^{(3a)}+\eta & \equiv-\frac{1}{\eta
^{2}}\phi_{k}^{(-a)}\bar{\phi}_{k}^{(a)}\phi_{k}^{(3a)}\equiv\phi_{k}^{(-a)}-\frac{1}{\eta}
\bar{\phi}_{k}^{(a)}\phi_{k}^{(3a)}-\eta\\
 & \equiv\phi_{k}^{(-a)}+\bar{\phi}_{k}^{(a)}+\phi_{k}^{(3a)}
\end{split}
\end{equation}
Note that for 1 kink and 1 anti-kink algebra is trivial because there are only 2 way we can
combine them mathematically according to ansatzs we have mentioned earlier.

But for 2 kinks and 1 anti-kink we can see there are 4 equivalent ways to combine using basic
algorithms(using product and additive ansatz). But note that apart from 2nd and 4th ansatz which
is trivial to write, 1st and 3rd way of writing is little non-trivial. And important thing to
notice is that changing relative sign between ways of first gluing kink anti-kink through product
then glu right kink through addition and first anti-kink kink through product then glu leftmost
kink with them through addition.

In more appropriate way is that gluing a left kink with right anti-kink and gluing left 
anti-kink with right kink is differ upto by a negative sign, therefore gluing operation is 
also not commutative through both addition and product operation. 

But both the operations are associative.

If we denote kinks and anti-kinks by their topological charge $+1$ and $-1$ respectively(there
is $0$ topological charge which denote a configuration whose asymptotic vacua are same) then
product operation $\otimes$(upto overall multiplicative factor) of gluing is defined in 
following way
\begin{equation}
\begin{split}
(+1)\otimes(-1) & =-((-1)\otimes(+1))\\
((+1)\otimes(-1))\otimes(+1) & =(+1)\otimes((-1)\otimes(+1))
\end{split}
\end{equation}
And if we denote addition operation of gluing by $\oplus$ (upto addition and subtraction 
by real number) then
\begin{equation}
\begin{split}
(+1)\oplus(-1) & =-((-1)\oplus(+1))\\
((+1)\oplus(-1))\oplus(+1) & =(+1)\oplus((-1)\oplus(+1))
\end{split}
\end{equation}
And according to earlier discussion
\begin{equation}
\begin{split}
((+1)\otimes(-1))\oplus(+1) & =(+1)\oplus(-((-1)\otimes(+1)))\\
(+1)\oplus(-1) & =(+1)\otimes(-1)\\
(-1)\oplus(+1) & =-((-1)\otimes(+1))
\end{split}
\end{equation}
(Note that rules in eq.(5.4) proves associativity of addition).

Similarly we have other relations in this algebra which are as fundamental as ones
are in above
\begin{equation}
\begin{split}
-((+1)\oplus(-1))\otimes(+1) & =(+1)\otimes((-1)\oplus(+1))\\
((-1)\oplus(+1))\otimes(-1) & =-(-1)\otimes((+1)\oplus(-1))\\
-((-1)\otimes(+1))\oplus(-1) & =(-1)\oplus((+1)\otimes(-1))
\end{split}
\end{equation}
Apart from $+1,-1$ other $\pm$ sign denotes actual signs used in addition algebra are used 
to glu the kink and anti-kink configurations with appropriate sign.

As we increase the number of kinks and anti-kinks such equivalent expressions are increasing 
drastically, for example for 2 kinks and 2 anti-kinks followings are few equivalent way of
writing a configuration whose left most part is kink
\begin{equation}
\begin{split}
\frac{1}{\eta^{3}}\phi_{k}^{(-a)}\bar{\phi}_{k}^{(a)}\phi_{k}^{(3a)}\bar{\phi}_{k}^{(5a)} & 
\equiv\phi_{k}^{(-a)}+\bar{\phi}_{k}^{(a)}+\phi_{k}^{(3a)}+\bar{\phi}_{k}^{(5a)}-\eta\\
 & \equiv\frac{1}{\eta^{2}}\phi_{k}^{(-a)}\bar{\phi}_{k}^{(a)}\phi_{k}^{(3a)}+\bar{\phi}_{k}
^{(5a)}-\eta\\
 & \equiv\frac{1}{\eta}\left(\frac{1}{\eta}\phi_{k}^{(-a)}\bar{\phi}_{k}^{(a)}+\phi_{k}^{(3a)}
+\eta\right)\bar{\phi}_{k}^{(5a)}\\
 & \equiv\frac{1}{\eta}\phi_{k}^{(-a)}\bar{\phi}_{k}^{(a)}+\phi_{k}^{(3a)}
+\eta+\bar{\phi}_{k}^{(5a)}-\eta\\
 & \equiv\phi_{k}^{(-a)}-\frac{1}{\eta}\bar{\phi}_{k}^{(a)}\phi_{k}^{(3a)}+\bar{\phi}_{k}
^{(5a)}-2\eta\\
 & \equiv\vdots 
\end{split}
\end{equation}
Above equivalent configurations can be obtained using our algebra which is as follows
\begin{equation}
\begin{split}
(+1)\otimes(-1)\otimes(+1)\otimes(-1) & \equiv((+1)\oplus(-1))\oplus((+1)\oplus(-1))\\
 & \equiv(((+1)\oplus(-1))\oplus(+1))\oplus(-1)\\
 & \equiv(((+1)\otimes(-1))\otimes(+1))\oplus(-1)\\
 & \equiv(((+1)\otimes(-1))\oplus(+1))\otimes(-1)\\
 & \equiv((+1)\otimes(-1))\oplus(+1)\oplus(-1)\\
 & \equiv(+1)\oplus(-((-1)\otimes(+1)))\oplus(-1)\\
 & \equiv\vdots
\end{split}
\end{equation}
Note that except 2nd line in above rest of line give respective configurations as we have
mentioned for 2 kinks and 2 anti-kink before above equation.

Before jump into conclusion part of this article, we want to clarify a doubt which 
is following.

Note that we have written the operation $(+1)\oplus(-1)=(+1)\otimes(-1)$ but if one
goes through following operations it seems that he/she would find inconsistent answer
\begin{equation}
(+1)\oplus(-1)\rightarrow-((-1)\oplus(+1))\rightarrow(-1)\otimes(+1)=-((+1)
\otimes(-1))
\end{equation} 
but this is actually wrong. So now the question is what went wrong?
The problem lies in the fact that in the fundamental did not change the order of
kink and anti-kink in more than once since in this case we swap the position of
them twice there is an extra $-$ sign arises.But swapping is only possible when 
we do it for every kink, anti-kink pair and in this case operation contains
1 kinks and 1 anti-kinks. But we consider total odd number of kinks, anti-kinks
then such swapping is not doable. We can think of it as non-commutativity in 
operational paths. 

\section{Conclusion}
Multi-kink configurations are found in many models like in \cite{deBrito:2014pqa} and their
collisions studied in \cite{Marjaneh:2017mko}. They can be used for transport process mentioned
in \cite{article}, \cite{PhysRevLett.119.153602} because of their approximate time dependent 
solution with long stability. Seeking explicit solutions for nonlinear evolution equations, 
by using different numerous methods, plays a major role in mathematical physics and becomes 
one of the most active areas of research investigation for mathematicians 
and physicists. In order to understand the physical mechanism of phenomena in nature, described 
by nonlinear PDEs, exact solutions of the nonlinear PDEs(partial differential equations) must 
be explored but often it is difficult to find an exact solution of such PDEs.  

In this article we follow slightly different approach and show how to construct approximate 
solutions by gluing basic building blocks which are kinks and anti-kinks. And then study their 
properties in terms of stabilities and shown indeed particle like properties of individual 
kinks and anti-kinks. By doing that we have shown that at fundamental level gluing kinks and 
anti-kinks in two different ways are equivalent. So we are basically looking at the collective 
behaviour of many individual kinks and anti-kinks which give rise to a multi-kink configuration 
although it is not long lasting in terms of stability issue. Then we went ahead and defined an 
algebra to give us the equivalent ways of constructing such multi-kink configurations. 
\section{Acknowledgement}
Author would like to thank CSIR to support this work through JRF fellowship.

\bibliographystyle{unsrt}
\nocite{*}
\bibliography{jhepexample}

\end{document}